\documentclass[review]{elsarticle}

\usepackage[margin=2.5cm]{geometry}
\usepackage{mathtools}
\usepackage{amsmath}
\usepackage{amssymb}
\usepackage{amsthm}
\usepackage{blkarray}
\usepackage{float}
\usepackage{graphicx}
\usepackage{multirow}
\usepackage{xcolor}
\usepackage{graphicx}
\usepackage{wrapfig}
\usepackage[utf8]{inputenc}
\usepackage[T1]{fontenc}
\usepackage{nomencl}
\usepackage{tabularx}
\usepackage{tabularx}
\usepackage{graphicx}
\usepackage{booktabs}
\usepackage{hyperref}
\usepackage{cancel}
\usepackage{lineno,hyperref}
\begin{document}

\begin{frontmatter}
    \title{
    		Numerical scheme for solving the nonuniformly forced 
    		cubic and quintic Swift-Hohenberg equations
    		strictly respecting the Lyapunov functional
    	   }
    \author[mymainaddress]{D. L. Coelho\corref{mycorrespondingauthor}}
    \cortext[mycorrespondingauthor]{Corresponding author. Tel.: +55 (21) 2332-4733}
    \ead{coelho.daniel@posgraduacao.uerj.br}
    \ead{danielcoelho.uerj@gmail.com}
    \author[mysecondaryaddress]{E. Vitral}
    \author[mymainaddress]{J. Pontes}
    \author[mymainaddress]{N. Mangiavacchi}
    \address[mymainaddress]{GESAR Group/UERJ - State University of
     Rio de Janeiro, 20940-903 Rua Fonseca Teles~121, Rio de Janeiro,
     RJ, Brazil}
    \address[mysecondaryaddress]{Department of Mechanical Engineering, University of Nevada, 1664 N. Virginia St. (0312), Reno, NV 9557-0312, U.S.A. }
    
    \begin{abstract}
    	Computational modeling of pattern formation in nonequilibrium systems
    	is a fundamental tool for studying complex phenomena in biology,
    	chemistry, materials and engineering sciences. The pursuit for
    	theoretical descriptions of some among those physical problems led to
    	the Swift-Hohenberg equation (SH3) which describes pattern selection
    	in the vicinity of instabilities. A finite differences scheme, known
    	as Stabilizing Correction (Christov \& Pontes; 2001 DOI:
    	10.1016/S0895-7177(01)00151-0), developed to integrate the cubic
    	Swift-Hohenberg equation in two dimensions, is reviewed and extended
    	in the present paper. The original scheme features Generalized
    	Dirichlet boundary conditions (GDBC), forcings with a spatial ramp of
    	the control parameter, strict implementation of the associated
    	Lyapunov functional, and second order representation of all
    	derivatives. We now extend these results by including periodic
    	boundary conditions (PBC), forcings with Gaussian distributions of
    	the control parameter and the quintic Swift-Hohenberg (SH35) model.
    	The present scheme also features a strict implementation of the 
    	functional for all test cases. A code verification was accomplished,
    	showing unconditional stability, along with second order
    	accuracy in both time and space. Test cases confirmed the monotonic
    	decay of the Lyapunov functional and all numerical experiments
    	exhibit the main physical features: highly nonlinear behaviour,
    	wavelength filter and competition between bulk and boundary effects.
    \end{abstract}
    \begin{keyword}
    Pattern Formation \sep  Nonlinear Systems \sep  
    Swift-Hohenberg Equation \sep  Finite Difference Methods
    \end{keyword}
\end{frontmatter}

\section{\label{intro}Introduction}
	Understanding the selection and orientation mechanisms of spatial
	structures, their symmetries and instabilities is a major
	theme of theoretical and experimental research in
	modeling of self-organization phenomena. Nowadays, spatiotemporal
	organization is recognized to be related to several technological
	problems in chemistry~\cite{yang2004stable}, nonlinear
	optics~\cite{newell2008}, and materials
	science~\cite{walgraefliv,ghoniem2008instabilities,provatas}. For
	computational material science and engineering a long-term goal is to
	adapt these models and appropriate numerical techniques to multiple scales and
	increasingly larger scale systems in order to achieve predictive
	capabilities~\cite{plapp2000multiscale,provatas2005multiscale,
	ghosh2017influence,ghosh2019influence}. 

	The pursuit for understanding self-organization in transport phenomena
	problems led to the Swift-Hohenberg (SH)~equation~\cite{Swift-1977}.
	It is a widely accepted model for describing pattern formation in
	many physical systems presenting symmetry breaking instabilities,
	such as the emergence of convection rolls in a thin layer of fluid
	heated from below. It first appeared in the framework of B\'enard
	thermal convection between two ``infinite'' horizontal surfaces with
	distinct temperatures, where J. Swift and P.
	Hohenberg~\cite{Swift-1977} performed a reduction of the full
	dynamics, led by the Boussinesq equations, into a dynamics governed by the most relevant slow modes represented by an order parameter. Such reduction can also be carried out for reaction-diffusion
	systems where a similar model is obtained with an additional
	quadratic nonlinearity
	(SH23)~\cite{walgraefliv,pontes2006dislocation}. Later, an
	extension of the original Swift-Hohenberg equation (SH3) was proposed
	with the introduction of a destabilizing cubic term and a quintic
	one. The resulting quintic SH equation (SH35) admits the coexistence
	of stable uniform and structured solutions,
	leading to an influx of studies on localized structures and snaking through the SH equation~\cite{sakaguchi1996,avitabile2010snake}.

	Roughly speaking, the SH equation has three basic pattern-forming
	mechanisms, related to each term of the equation. The first is a control parameter (e.g. distance to temperature threshold in the Rayleigh-Bénard convection) in the linear part of the equation, which will be addressed to as a
	forcing parameter. It represents how far from the onset of the
	instability the system is. The second is the wavelength filter that contains
	all the equation spatial derivatives, and leads to wavelength selection in the resulting patterns. The third one is the nonlinear part of the equation,  whose terms promote interactions between modes involved in the dynamics. For example, in the
	SH3 the cubic term is responsible for saturating the linear growth.
	Meanwhile, in the SH35 the cubic term destabilizes the dynamics leaving the
	quintic one responsible for the saturation. Some generalizations of theses mechanisms appear in literature, such as theoretical works
	exploring nonuniform forcings in the Rayleigh-Bénard 
	convection~\cite{walton1982onset,walton1983onset}.

	The SH equation admits a potential, formally known as the Lyapunov functional, from which the dynamical equation governing the evolution of the order parameter can be derived. In recent
	years, this potential nature has been exploited in condensed matter
	as a physical feature associated with the system's
	free energy, used to describe patterning in materials.
	In this context, the SH dynamics is part of the phase-field
	theory, which was developed from statistical mechanics principles,
	and whose goal is to obtain governing equations for microstructure
	evolution; it connects thus thermodynamic and kinetic properties with
	microstructure via a mathematical
	formalism~\cite{provatas,provatas2005multiscale,elder2002modeling,elder2004modeling}.
	The phase-field theory is a descendant of the van der Waals,
	Cahn-Hilliard and Landau type classical field theoretical approaches,
	originating from a single order parameter gradient theory of
	Langer~\cite{langer1978theory,langer1980instabilities}. In this
	theory, the local state of matter is characterized by an
	order parameter $\psi(\mathbf{x},t)$, the phase-field
	variable, which monitors the  transition between phases of distinct
	order of property. It can represent the structural order parameter
	that measures local crystallinity, the composition of a phase, the
	degree of a molecular ordering, among others. In the case of the phase-field
	crystal (PFC) approach, a conserved dynamics is assumed for this
	order parameter. Although this formalism will not be addressed in
	this paper, many recent theoretical and numerical works~\cite{elsey2013simple,lee2016simple,li2017efficient} are interested on conserved SH forms for PFC applications 
	
	Naturally, these applications led to the development of a series of mathematical and numerical methods concerned with the study of the SH equation
	In the beginning of this endeavor,
	Greenside  \& Coughran~Jr. (1984) \cite{Greenside1984} achieved
	relevant results through a finite differences approach by using a
	backward Euler time-integration scheme with
	rigid~\cite{cross1982ingredients,cross1982boundary} and periodic
	boundary conditions (PBC). Those types of boundary conditions were
	also explored by Manneville~(1990)\cite{Manneville1990} and
	Cross~(1994)\cite{Cross1993}. These works considered uniform control
	parameters. Other authors employed nonuniform distributions (ramps
	and Gaussians) for the control parameter and presented very
	interesting results as well, displaying rich competition between bulk
	and boundary effects. In particular, ramped control parameters mostly appears in
	the context of rigid boundaries~\cite{C.I-1997,pontes2008}, which
	will be addressed here as Generalized Dirichlet boundary conditions
	(GDBC), following~\cite{C.I-1997,C.I-2002,pontes2008}.

	Besides finite differences, spectral methods have become largely employed due to their highly spatial
	accurate solutions~\cite{lee2019energy,vitral2019role,wang2020fast}. They
	usually treat the nonlinear terms explicitly in the physical space, and
	therefore are called pseudo-spectral methods. Fully explicit schemes lead to
	severe stability restrictions on the time step, due to the the fourth-order
	nonlinear partial differential equation originated from the non-conserved
	dynamics. In the case of the PFC applications, the conserved dynamics is governed by a six-order nonlinear partial differential equation,
	which is even more restrictive for the time step selection. While PFC is not
	a subject of this work, our proposed scheme can be readily translated into a
	numerical framework for many PFC problems.

	The present article is a continuation of the work presented by
	Christov \& Pontes~(2002)\cite{C.I-2002}, where the authors employed
	a finite differences scheme 
	for solving
	the cubic Swift-Hohenberg equation (SH3) in two dimensions, with
	uniform forcing, GDBC, and strict implementation of the associated functional.
	This scheme, known as Stabilizing
	Correction~\cite{Douglas,Yanenko}, was originally introduced by Douglas \&
	Rachford in the context of the temperature equation, 
	and has also been adopted for solving 
	fourth-order nonlinear differential equations similar to SH, such as the Kuramoto-Sivashinsky equation~\cite{Vitral-2018}.
	Here, we expand the previous study on the SH equation~\cite{C.I-2002} for
	additional nonlinearities, boundary conditions and a spatially
	dependent control parameter. This includes the quintic SH35 with both GDBC and PBC, nonuniform
	Gaussian distributions of the control parameter in addition to ramps,
	and also a strict implementation of the associated Lyapunov
	functional. The scheme features a semi-implicit time splitting with
	second order representation of time and space derivatives. 
	It is pertinent to point out that operator splitting methods
	enable parallel code processing implementation, which allows faster
	simulations. The usage of  sparse matrices also plays a huge role in
	lowering computational cost in the simulations, and therefore it was
	implemented in the present work.

	A code verification is performed to confirm the unconditional
	stability of the scheme, along with its second order convergence in
	both time and space. Since nonlinear partial differential equations such as SH
	do not have easily accessible analytical solutions, the adopted method to verify the order of accuracy of the code
	is the method of manufactured solutions (MMS). It provides a
	convenient way of verifying the implementation of nonlinear numerical
	algorithms, since we can use  an artificial (manufactured) solution
	for this
	purpose~\cite{roache2002code,roy2005review,Vitral-2018}.

	This paper is organized as follows. Section~\ref{sec:mathmodel}
	briefly discusses the properties of the SH equation and the numerical
	scheme for solving the cubic and quintic versions of the equation
	with GDBC and PBC, nonuniform forcings, and the discrete
	implementation of the associated Lyapunov functional.
	Section~\ref{codevef} contains all verification procedures adopted.
	Section~\ref{sec:numerical experiments} presents the numerical
	experiments. The results are grouped in three parts; the first one
	showing the steady state patterns obtained in ten simulations with
	the SH3 equation; the second one showing six simulations with the
	SH35 and the last one with discussions. Section~\ref{sec:conclusions}
	contains the final remarks of the work.

%%%%%%%%%%%%%%%%%%%%%%%
\section{\label{sec:mathmodel}Mathematical modelling and numerical scheme}
% Methods
% \input{./sections/mathmodel.tex}

	The mathematical description of the dimensionless governing
	equations is briefly discussed in the start of this section.
	Next, the details of the numerical framework are discussed as an
	extension of Christov \& Pontes~(2002)\cite{C.I-2002} work. For both
	SH3 and SH35 discretizations, a finite difference semi-implicit
	scheme of second order accuracy in time and space is presented.
	Finally, the choices for the operator splitting method, spatial
	discretizations and mesh types are clarified.

\subsection{Governing equations}

	The SH equation has the so-called gradient dynamics, which means
	there is  a potential function of the order parameter
	$\psi\left({\bf x},t\right)$, (known as a Lyapunov functional),
	that has the property of decreasing monotonically during the
	evolution~\cite{C.I-2002}. This dynamic equation can be derived 
	from the $L^2$-gradient flow of the Lyapunov energy functional:
	\begin{align}
		\mathcal{F}[\psi]
		&\vcentcolon =
		\int_\Omega {d\mathbf{x}}\dfrac{1}{2}
		\left\{
			-{\epsilon(\mathbf{x})}\psi^2
			+
			{\alpha}
			\left[(q_0^2+\nabla^2)\psi\right]^2
			-
			\dfrac{\beta}{2}\psi^4+\dfrac{\gamma}{3}\psi^6
		\right\};
		\label{eq:LF1}
		\\
		\dfrac{\partial\mathcal{F}[\psi]}{\partial t}
		&=
		-\int_\Omega {d\mathbf{x}}
		\left(\dfrac{\partial\psi}{\partial t}\right)^2 \leq 0.
	\label{eq:dec_LF}
	\end{align}
	where $\Omega$ represents the domain whose size is commensurate with
	the length scales of the patterns. We consider a regular domain $
	\Omega:\{x\in [0,L_x]$, $y \in [0,L_y]\}$. Equation~\eqref{eq:dec_LF}
	denotes the  nonincreasing behavior of the Lyapunov functional,
	which monotonically  decreases until steady state is
	reached~\cite{C.I-2002}. By taking the variational derivative of 
	Eq.~\eqref{eq:LF1} in $L^2$ norm, the  Swift-Hohenberg
	equation is obtained:
\begin{equation}
\centering
\dfrac{\partial\psi}{\partial t}=
- \mu=
\epsilon(\mathbf{x})\psi-\alpha\left(q_0^2+\nabla^2\right)^2\psi+
\beta\psi^3-\gamma\psi^5 \,,
\label{SH}
\end{equation}
where $\mu = {\delta\mathcal{F}[\psi]}/{\delta\psi}$ is the chemical potential. As discussed by Christov \textit{et al.}~\cite{C.I-2002,C.I-1997},
	the appropriate boundary conditions for this system should consider
	only production and dissipation in the bulk and not on the boundary
	($\partial \Omega$). This can be achieved by assuming Generalized
	Dirichlet boundary conditions (GDBC), $v=\partial v/ \partial n=0$,
	where $n$ stands for the outward normal direction to the boundary.
	Periodic boundary conditions (PBC) will also be addressed in this
	work, since the absence of boundaries also leads boundary ($\partial
	\Omega$) integrals to vanish.

	We expand Eq.~\eqref{SH} by considering the following laplacian
	operator in cartesian coordinates: $\nabla^2\equiv{\partial^2}/
	{\partial x^2}+{\partial^2}/{\partial y^2}$, we have:

	\begin{equation}
		\begin{aligned}
			\dfrac{\partial\psi}{\partial t}   =  &\;
			\epsilon(\mathbf{x})\psi-\alpha q_0^4\psi
			-
			2\alpha q_0^2\nabla^2\psi-\alpha \nabla^4\psi
			+
			\beta\psi^3-\gamma\psi^5
			\\
			=  & \;
			\epsilon(\mathbf{x})\psi-\alpha q_0^4\psi
			-
			2\alpha q_0^2\dfrac{\partial^2\psi}{\partial x^2}
			-
			2\alpha q_0^2\dfrac{\partial^2\psi}{\partial y^2}
			-
			\alpha \dfrac{\partial^4\psi}{\partial x^4}
			-
			2\alpha \dfrac{\partial^4\psi}{\partial x^2 \partial y^2}
			-
			\alpha\dfrac{\partial^4\psi}{\partial y^4}
			+
			\beta\psi^3-\gamma\psi^5.
		\end{aligned}
		\label{SH2}
	\end{equation}

	\begin{table}[H]
		\centering
		\caption{\label{tab:parameters}Parameters assumed for the
		governing equations studied in this work.}
		\vspace*{10pt}
		\centering
		\setlength\tabcolsep{4.0 pt}
		\renewcommand{\arraystretch}{1.0}
		\begin{tabular}{@{}c c c c c c @{}}
			\toprule [{1.5pt}]
			\textbf{Equation}  & \textbf{Nonlinearity}   & $\alpha$  & $\beta$ &
			$\gamma$ & $q_0$
			\\
			\midrule
			SH3       & cubic          & 1    & -1 & 0   & 1
			\\[0.2mm]
			SH35      & quintic  & 1   & 1   & 1   & 1
			\\
			\bottomrule [{1.5pt}]
		\end{tabular}
		\label{tab:param}
	\end{table}

	By using the parameters and values of Table~\ref{tab:param} in
	Eq.~\eqref{SH}, we define the forms of the SH3 and SH35. The
	discretization of these governing equations is described in the
	following subsection.

\subsection{The target scheme}
	In order to construct a Crank-Nicolson second order in time
	numerical scheme, we adopt the following representation proposed by
	Christov and Pontes (2002)~\cite{C.I-2002} for the time derivative
	of Eq.~\eqref{SH}, where the RHS is evaluated at the middle of the
	time step $\Delta t$. The updated scheme, now including a quintic
	term takes the form: 
{
	\begin{align}
		\dfrac{\psi^{n+1}-\psi^{n}}{\Delta t}
		\,=\,
		-\mu^{n+1/2}
		\,=\, &
		\left[
			\epsilon(\mathbf{x})-\alpha q_0^4
			-
			2\alpha q_0^2\dfrac{\partial^2}{\partial x^2}
			-
			2\alpha q_0^2\dfrac{\partial^2}{\partial y^2}
			-
			\alpha \dfrac{\partial^4}{\partial x^4}
			-
			2\alpha \dfrac{\partial^4}{\partial x^2\partial y^2}
			-
			\alpha\dfrac{\partial^4}{\partial y^4}
		\right.
		\nonumber
		\\ &
		\left.
			+\beta\dfrac{\left(\psi^{n+1}\right)^2
			+
			\left(\psi^n\right)^2}{2}
			-
			\gamma\dfrac{\left(\psi^{n+1}\right)^4
			+
			\left(\psi^n\right)^4}{2}
		\right]
		\left(
			\dfrac{\psi^{n+1}+\psi^{n}}{2}
		\right).
		\label{eq:numericalscheme}
	\end{align}
}

	The superscript $(n+1)$ refers to the next time to be evaluated and
	$n$, to the current one. The parameter values for $\alpha$, $\beta$,
	$\gamma$ and $q_0$ are chosen according to Table~\ref{tab:param}.
	The RHS terms of Eq.~\eqref{eq:numericalscheme} are grouped in three parts, the
	first and the second ones containing the semi-implicit operators
	$\Lambda^{n+1/2}_x$ and $\Lambda^{n+1/2}_y$ that act on the variable
	$\left(\psi^{n+1}+\psi^n\right)/2$, and a function $f^{n+1/2}$, evaluated at
	the middle of the time step $\Delta t$. This function will contain explicit
	terms only, in the final discrete form of Eq.~\eqref{eq:numericalscheme}. Space
	derivatives are represented by centered second order formul{\ae}. Implicit
	terms are chosen having in mind to construct negative definite operators that
	will assure the stability of the scheme. The factor $1/2$ multiplying
	$\left(\psi^{n+1}+\psi^n\right)$ in the RHS of the above equation is included
	in the operators
	$\Lambda_{x}^{n+1/2}$ and $ \Lambda_{y}^{n+1/2}$, leading to the
	following target scheme:
	\begin{equation}
		\dfrac{\psi^{n+1}-\psi^{n}}{\Delta t}
		= 
		\left(\Lambda_{x}^{n+1/2}+\Lambda_{y}^{n+1/2}\right)
		\left(\psi^{n+1}+\psi^{n}\right)+f^{n+1/2}.
		\label{TS1}
	\end{equation}
	For the SH3 equation, the operators $\Lambda_{x}^{n+1/2}$,
	$\Lambda_{y}^{n+1/2}$ and $f^{n+1/2}$ are defined as:
	\begin{equation}
	\begin{aligned}
		\Lambda_{x}^{n+1/2} = & \;
		\dfrac{1}{2}
		\left[
			-\alpha
			\left(
				\dfrac{\partial^4}{\partial x^4}+\dfrac{q_0^4}{2}
			\right)
			-
			\beta
			\dfrac{\left(\psi^{n+1}\right)^2+\left(\psi^n\right)^2}{2}
		\right];
		\\
		\Lambda_{y}^{n+1/2} = & \;
		\dfrac{1}{2}
		\left[
			-\alpha
			\left(
				\dfrac{\partial^4}{\partial y^4}+\dfrac{q_0^4}{2}
			\right)
			-
			\beta
			\dfrac{\left(\psi^{n+1}\right)^2+\left(\psi^n\right)^2}{2}
		\right];
		\\
		f^{n+1/2} = &  \;
		\dfrac{1}{2}
		\left[
			\epsilon(\mathbf{x})
			-\alpha
			\left(
				2q_0^2\dfrac{\partial^2}{\partial x^2}
				+
				2q_0^2\dfrac{\partial^2}{\partial y^2}
				+
				2\dfrac{\partial^4}{\partial x^2\partial y^2}
			\right)
		\right]
		\left(\psi^{n+1}+\psi^{n}\right),
		\label{eq:operators SH3}
	\end{aligned}
	\end{equation}
	and for the SH35:
	\begin{eqnarray}
	 \Lambda_{x}^{n+1/2} &=&
		\dfrac{1}{2}
		\left[
			-\alpha
			\left(
				\dfrac{\partial^4}{\partial x^4}+\dfrac{q_0^4}{2}
			\right)
			-
			\gamma
			\dfrac{\left(\psi^{n+1}\right)^4+\left(\psi^n\right)^4}{2}
		\right];
		\nonumber
		\\
		\Lambda_{y}^{n+1/2} &=&
		\dfrac{1}{2}
		\left[
			-\alpha
			\left(
				\dfrac{\partial^4}{\partial y^4}+\dfrac{q_0^4}{2}
			\right)-
			\gamma
			\dfrac{\left(\psi^{n+1}\right)^4+\left(\psi^n\right)^4}{2}
		\right];
		\nonumber
		\\
		f^{n+1/2} &=&
		\dfrac{1}{2}
		\left[
			\epsilon(\mathbf{x})
			-
			\alpha
			\left(
				2q_0^2\dfrac{\partial^2}{\partial x^2}
				+
				2q_0^2\dfrac{\partial^2}{\partial y^2}
				+
				2\dfrac{\partial^4}{\partial x^2\partial y^2}
			\right)
		\right.
		\nonumber
		\\
		&&
		\left.
			+
			\beta
			\dfrac{\left(\psi^{n+1}\right)^2+\left(\psi^n\right)^2}{2}
		\right]
		\left(\psi^{n+1}+\psi^{n}\right).
		\label{eq:operators SH35}
	\end{eqnarray}

\subsection{Internal iterations}

	Since the operators $\Lambda_{x}^{n+1/2}$, $\Lambda_{y}^{n+1/2}$ and the
	function $f^{n+1/2}$ in Eqs.~\ref{TS1}, \ref{eq:operators SH3} and
	\ref{eq:operators SH35} contain implicit terms due to the time discretization
	and linearization of the nonlinear terms, we do internal iterations. They are
	required to secure the approximation of the nonlinearities in the scheme
	(Eq.~\eqref{TS1}) at each time step. The iterations loop proceeds until
	convergence is attained by monitoring the $L_\infty$ norm. The internal
	iterations scheme reads:

	\begin{equation}
		\dfrac{\psi^{n+1,p+1}-\psi^{n}}{\Delta t}
		= 
		\left(\Lambda_{x}^{n+1/2,p}+\Lambda_{y}^{n+1/2,p}\right)
		\left(\psi^{n+1,p+1}+\psi^{n}\right)+f^{n+1/2,p},
		\label{TS}
	\end{equation}
	where the index $(p)$ refers to the internal iteration number. The
	superscript $(n+1,p+1)$ identifies the new iteration, while $(n)$ are
	the values of the previous time step. The superscript $(n+1)$ for
	the nonlinear term in the function $f^{n+1/2}$ will be replaced by
	$(n,p)$, which stands for the values obtained from the previous
	iteration. The operators $\Lambda_{x}^{n+1/2,p}$, $\Lambda_{y}^{n+1/2,p}$ function
	$f^{n+1/2,p}$ are redefined as follows, for the SH3 equation:

	\begin{eqnarray}
		\Lambda_{x}^{n+1/2,p} &=&
		\dfrac{1}{2}
		\left[
			-\alpha
			\left(
				\dfrac{\partial^4}{\partial x^4}+\dfrac{q_0^4}{2}
			\right)
			-
			\beta
			\dfrac{\left(\psi^{n+1,p}\right)^2+\left(\psi^n\right)^2}{2}
		\right],
		\nonumber
		\\
		\Lambda_{y}^{n+1/2,p} &=&
		\dfrac{1}{2}
		\left[
			-\alpha
			\left(
				\dfrac{\partial^4}{\partial y^4}+\dfrac{q_0^4}{2}
			\right)
			-
			\beta
			\dfrac{\left(\psi^{n+1,p}\right)^2+\left(\psi^n\right)^2}{2}
		\right],
		\nonumber
		\\
		f^{n+1/2,p} &=&
		\dfrac{1}{2}
		\left[
			\epsilon(\mathbf{x})
			-\alpha
			\left(
				2q_0^2\dfrac{\partial^2}{\partial x^2}
				+
				2q_0^2\dfrac{\partial^2}{\partial y^2}
				+
				2\dfrac{\partial^4}{\partial x^2\partial y^2}
			\right)
		\right]
		\left(\psi^{n+1,p}+\psi^{n}\right),
	\end{eqnarray}
	and for the SH35:

	\begin{eqnarray}
	 \Lambda_{x}^{n+1/2,p} &=&
		\dfrac{1}{2}
		\left[
			-\alpha
			\left(
				\dfrac{\partial^4}{\partial x^4}+\dfrac{q_0^4}{2}
			\right)
			-
			\gamma
			\dfrac{\left(\psi^{n+1,p}\right)^4+\left(\psi^n\right)^4}{2}
		\right],
		\nonumber
		\\
		\Lambda_{y}^{n+1/2,p} &=&
		\dfrac{1}{2}
		\left[
			-\alpha
			\left(
				\dfrac{\partial^4}{\partial y^4}+\dfrac{q_0^4}{2}
			\right)-
			\gamma
			\dfrac{\left(\psi^{n+1,p}\right)^4+\left(\psi^n\right)^4}{2}
		\right],
		\nonumber
		\\
		f^{n+1/2,p} &=&
		\dfrac{1}{2}
		\left[
			\epsilon(\mathbf{x})
			-
			\alpha
			\left(
				2q_0^2\dfrac{\partial^2}{\partial x^2}
				+
				2q_0^2\dfrac{\partial^2}{\partial y^2}
				+
				2\dfrac{\partial^4}{\partial x^2\partial y^2}
			\right)
		\right.
		\nonumber
		\\
		&&
		\left.
			+
			\beta
			\dfrac{\left(\psi^{n+1,p}\right)^2+\left(\psi^n\right)^2}{2}
		\right]
		\left(\psi^{n+1,p}+\psi^{n}\right).
	\end{eqnarray}
	\par 
	The iterations proceed until the following criterion for the
	$L_\infty$ norm is satisfied with $\delta=1.0 \times 10^{-8}$:
	\begin{equation}
		L_\infty = \dfrac{\max\mid \psi^{n+1,p+1}-\psi^{n+1,p}\mid}
		{\max\mid \psi^{n+1,p+1}\mid}\leq\delta;
	\end{equation}
	\noindent
	so that the last iteration gives the sought function $\psi$ in the new time
	$\psi^{n+1} \equiv \psi^{n+1,p+1}$.
\subsection{The splitting scheme}
\smallskip 

	Although employing sparse matrices for the operators, computationally solving
	Eq.~\eqref{TS} still represents a costly procedure. In order to reduce such
	computational effort and errors (from discretization and floating-point
	operations), the operators splitting method was adopted. The splitting of
	Eq.~\eqref{TS1} is made according to the Douglas second scheme (also known as
	scheme of stabilizing correction, shown by \cite{C.I-1997},\cite{C.I-2002}),
	and is briefly reviewed. The following equations represent a consistent
	approximation of the original scheme:
	
	\begin{align}
		\dfrac{\widetilde{\psi}-\psi^{n}}{\Delta t}
		=  & \;
		\Lambda_{x}^{n+1/2,p}\widetilde{\psi}+\Lambda_{y}^{n+1/2,p}\psi^{n}
		+
		f^{n+1/2,p}+(\Lambda_{x}^{n+1/2,p}+\Lambda_{y}^{n+1/2,p})\psi^{n},
		\label{SS1}
		\\
		\dfrac{\psi^{n+1,p+1}-\widetilde{\psi}}{\Delta t}
		=  & \;
		\Lambda_{y}^{n+1/2,p}(\psi^{n+1,p+1}-\psi^{n}),
	\label{SS2}
	\end{align}
	where $\widetilde\psi$ is an intermediary estimation of
	$\psi$ at the new time step. In order
	to show that the splitting represents the original scheme, we
	rewrite Eqs.~\ref{SS1} and~\ref{SS2} in the form:
	\begin{align}
		\left(E-\Delta t \Lambda_{x}^{n+1/2,p}\right)\widetilde\psi
		=  &
		\left(E+\Delta t \Lambda_{x}^{n+1/2,p}\right)\psi^{n}
		+
		2\Delta t\Lambda_{y}^{n+1/2,p}\psi^{n}+\Delta t f^{n+1/2,p},
		\label{Ss1}
		\\
		\left(E-\Delta t \Lambda_{y}^{n+1/2,p}\right)\psi^{n+1,p+1}
		=  & \;
		\widetilde\psi-\Delta t\Lambda_{y}^{n+1/2,p}\psi^{n},
		\label{Ss2}
	\end{align}

	\noindent where $E$ is the unity operator. Rearranging these
	equations, the intermediate variable $\widetilde\psi$ is eliminated
	and the result may be rewritten as:

	\begin{align}
		\left(E-\Delta t \Lambda_{x}^{n+1/2,p}\right)
		\left(E-\Delta t \Lambda_{y}^{n+1/2,p}\right)\psi^{n+1,p+1}
		=  &
		\left(E+\Delta t \Lambda_{x}^{n+1/2,p}\right)\psi^{n}
		+
		2\Delta t\Lambda_{y}^{n+1/2,p}\psi^{n}+
		\nonumber
		\\
		& + \Delta t f^{n+1/2,p} -
		\left(E-\Delta t \Lambda_{x}^{n+1/2,p}\right)
		\Delta t\Lambda_{y}^{n+1/2,p}\psi^{n}.
	\label{Ss244}
	\end{align}

	\par This result may be rewritten as:

	\begin{align}
		\left(E+\Delta t^2 \Lambda_{x}^{n+1/2,p}\Lambda_{y}^{n+1/2,p}\right)
		\dfrac{\psi^{n+1,p+1}-\psi^{n}}{\Delta t}
		=  & \;
		(\Lambda_{x}^{n+1/2,p}+\Lambda_{y}^{n+1/2,p})
		(\psi^{n+1,p+1}+\psi^{n})+f^{n+1/2,p},
		\label{Ss242424}
	\end{align}
	where $E$ is the unity operator. A comparison with Eq.~\eqref{TS1}
	shows that Eq.~\eqref{Ss242424} is actually equivalent to the former
	except by the positive definite operator having a norm greater than
	one:
	\begin{equation}
		B \equiv E+\Delta t^2\Lambda_{x}^{n+1/2,p}\Lambda_{y}^{n+1/2,p}
		=
		E+\mathcal{O}(\Delta t^2),
	\end{equation}
	which acts on the term $(\psi^{n+1,p+1}-\psi^{n})/{\Delta t}$. This
	means that this operator does not change the steady state solution,
	and since $||B|| > 1$, the scheme given by
	Eqs.~\ref{SS1}, and~\ref{SS2} is more stable than the target one
	(Eq.~\eqref{TS1}).
\subsection{Spatial discretization and boundary conditions}
\begin{figure}[htb!]
		\centering
		\includegraphics[width=0.95\textwidth]{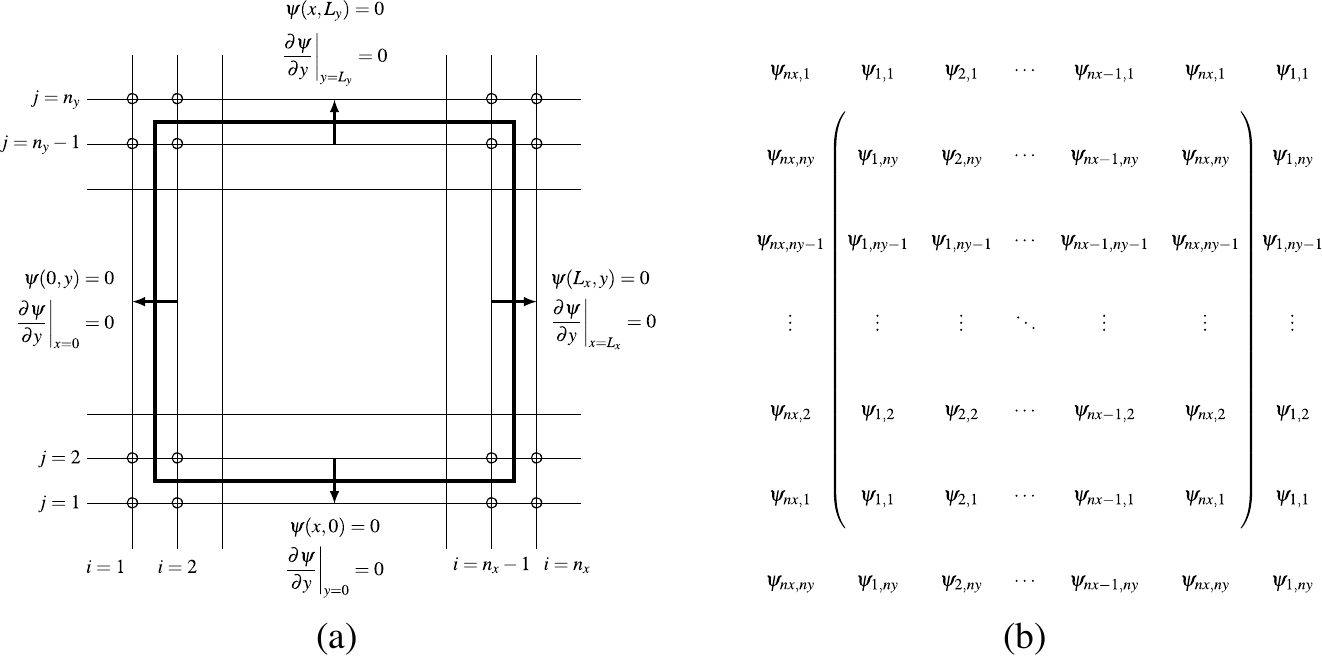}
		\caption{\label{fig:grids} The grids and boundary conditions used in this
		work. (a) The ``\textit{Staggered}'' grid. In the case of GDBC, values of
		$\psi$ in the first two lines and in the two last ones, and also in the first
		two columns and in the last two ones are assigned to zero. (b) The periodic
		domain.}
	\end{figure}

	We solve numerically the SH equation with GDBC and PBC. For
	the case of rigid boundary conditions we adopt a staggered grid, as
	illustrated in Fig.~\ref{fig:grids}. Consider a staggered mesh in
	both spatial directions, namely
	\begin{equation*}
		x_i=-\dfrac{\Delta x}{2}+i\,\Delta x,
		\quad
		\Delta x\equiv\dfrac{L_x}{n_x-2},
		\quad
		y_j=-\dfrac{\Delta y}{2}+j\,\Delta y,
		\quad
		\Delta y\equiv\dfrac{L_y}{n_y -2},
	\end{equation*}
	where $n_x$ and $n_y$ are the number of points in $x$-and
	$y$-directions, respectively. The mesh pattern is shown in
	Fig.~\ref{fig:grids}. Let $\psi_{i,j}$ be an arbitrary set function
	defined on the above described mesh. We confine ourselves to the
	case of constant coefficients. The PBC, which is less restrictive
	than the GDBC, borrows the crystallography concept of unit cell,
	which is a repeating pattern representative  of the material. That is,
	using PBC we have a domain that would also act as a unit cell for a
	infinite two-dimensional surface, with cells being periodically
	repeated all around the bidimensional domain.

	Then the simplest second order symmetric difference approximations
	of the differential operators are obtained by making a Taylor
	development of a function $\psi$ at the points of a uniform grid.
	Considering the function $\psi$ in a bidimensional domain, we can
	define the derivatives by truncating the Taylor expansion and
	arranging the equations. 

	Second derivatives with second order accuracy can be written as:
	\begin{equation}
	\begin{aligned}
		\dfrac{\partial^2\psi_{i,j}}{\partial x^2} =
		&  \;&
		\dfrac{\psi_{i,j-1}-2\psi_{i,j}+\psi_{i,j+1}}{\Delta x^2};
		&\hspace*{10pt}
		{\hspace{10pt}{\dfrac{\partial^2\psi_{i,j}}{\partial y^2}=}}
		&\;& {{\dfrac{\psi_{i-1,j}-2\psi_{i,j}+
		\psi_{i+1,j}}{\Delta y^2}}}.
	\end{aligned}
	\end{equation}
	And fourth derivatives with second order accuracy are written as:
	\begin{align*}
		\dfrac{\partial^4\psi_{i,j}}{\partial x^4}
		=  &
		\dfrac{\psi_{i-2,j}-4\psi_{i-1,j}+6\psi_{i,j}-4\psi_{i+1,j}+
		\psi_{i+2,j}}{\Delta x^4},
		\\
		\dfrac{\partial^4\psi_{i,j}}{\partial y^4}
		=  &
		\dfrac{\psi_{i,j-2}-4\psi_{i,j-1}+6\psi_{i,j}-
		4\psi_{i,j+1}+\psi_{i,j+2}}{\Delta y^4},
		\\
		\dfrac{\partial^4\psi_{i,j}}{\partial x^2\partial y^2}
		=  &
		\dfrac{1}{\Delta x^2\Delta y^2}
		\Bigg(
			\psi_{i-1,j-1}-2\psi_{i,j-1}+\psi_{i+1,j-1}
			-
			2\psi_{i-1,j}+4\psi_{i,j}-2\psi_{i+1,j}+
			\\
			&
			+\psi_{i-1,j+1}-2\psi_{i,j+1}+\psi_{i+1,j+1}
		\Bigg).
\end{align*}

    Standard second order representations of spatial derivatives are
	adopted in uniform and structured grids. In order to verify the
	correctness of the implementation, we conducted convergence tests
	using the method of manufactured solutions (MMS). The results can be
	found in Sec.~\ref{codevef}.

\section{\label{codevef}Code verification}

	One important issue concerning the simulations is the time and mesh
	size selection, such that we seek reasonable choices inside the
	stable region of the proposed numerical scheme. The governing
	equation~\eqref{SH} can be rewritten in the form
	$P\psi(\mathbf{x},t)=0$, where $P$ is an operator containing all the
	partial derivatives and terms acting on the order parameter
	$\psi(\mathbf{x},t)$. The consistency of the numerical scheme can be
	easily verified since $P\psi(\mathbf{x},t)-P_{\Delta t,\Delta
	x,\Delta y}\psi(\mathbf{x},t) \longrightarrow 0$ as $\Delta t,\Delta
	x,\Delta y \longrightarrow 0$, where $P_{\Delta t,\Delta x,\Delta y}$
	is the finite difference discretization of $P$. In this section,
	scheme stability, free energy functional decay and convergence tests
	were performed to verify the code implementation.

\subsection{Scheme stability}

\begin{figure}[htb!]
		\centering
		\includegraphics[width=0.9\textwidth]{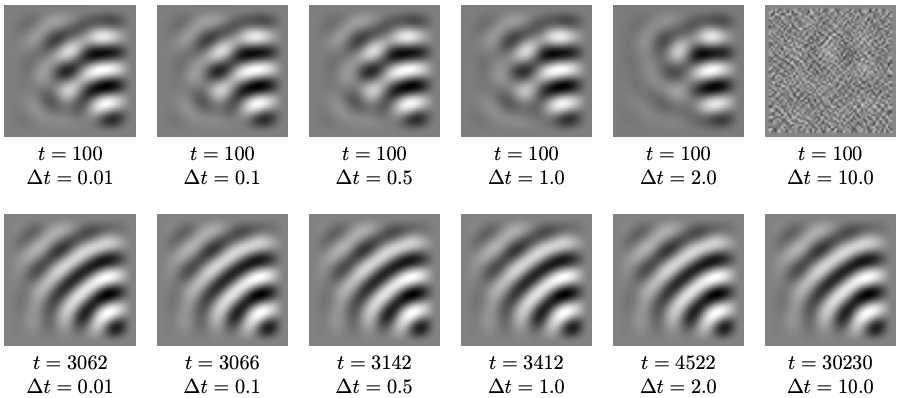}
		\caption{\label{fig:stabilitypics1} Pattern developed with the SH3
		model, GDBC, forced with a spatial ramp of the control parameter given
		by $0.0\leq\epsilon(\mathbf{x})\leq0.2$, and six different time steps.
		Top row: transient states at $t=100$. Bottom: the steady state for the
		six time steps. All tests started from the exactly same initial
		condition (pseudo-randomly generated) for a $64 \times 64$ nodes
		domain. Note that the use of larger time steps results in smaller
		number of required steps to attain the same ``time'' ($t=100$ for
		instance). Errors with larger time steps result in delay in the
		emergence of the pattern and thus later steady states is expected.} 
	\end{figure}

\begin{figure}[htb!]
		\centering
		\includegraphics[width=0.9\textwidth]{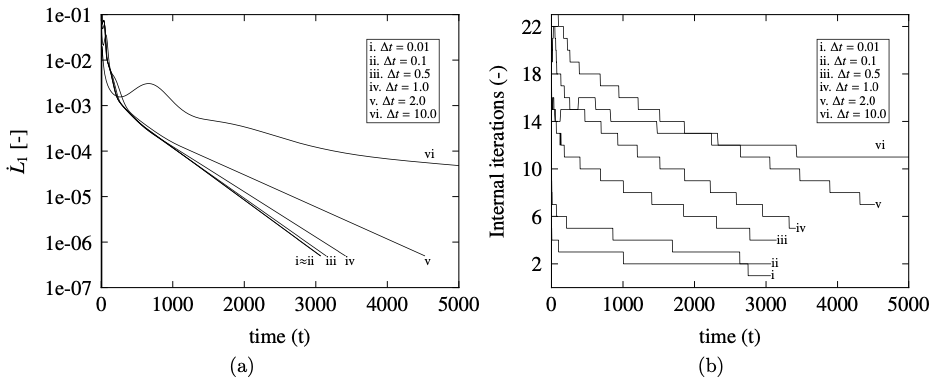}
		\caption{\label{fig:stabilitypics2}Scheme stability analysis from
		the numerical integration of the Swift-Hohenberg equation. (a)
		$\dot{L}_1$ and (b) internal iterations are shown as a function of
		time $t$ for a $64 \times 64$ nodes domain for all six tests are
		presented in Fig.~\ref{fig:stabilitypics1}. The same steady pattern is
		reached for all six analyzed cases independently of the time step. The
		simulations started from the same random initial conditions.}
	\end{figure}

	Here, following Christov and Pontes (2001)\cite{C.I-2002}, we study
	the effect of the time step in the structure evolution by assessing
	the rate of change in time of the pattern during the simulation. We
	do this by monitoring $\dot{L}_1$, the relative norm rate of change
	defined as:
	
	\begin{equation}
		\dot{L}_1\;=\dfrac{1}{\Delta t}\;\left(
		\frac{\sum\limits_{i=1}^{n_x}\sum\limits_{j=1}^{n_y}\mid \psi_{i,j}^{n+1}-\psi_{i,j}^{n}\mid}
		{\sum\limits_{i=1}^{n_x}\sum\limits_{j=1}^{n_y}\mid \psi_{i,j}^{n+1}\mid}\right),
		\label{eq:L1}
	\end{equation}
	which roughly corresponds to the ratio between the spatial average of
	the modulus of time derivative ${\partial \psi}/{\partial t}$ and the
	spatial average of the modulus of the function itself. The
	calculations begin from a random initial condition and proceeded
	until $\dot{L}_1\leq 5 \times 10^{-7}$, which is our criterion for
	reaching the steady state. Following this implementation,
	Fig.~\ref{fig:stabilitypics1} shows the system state at $t=100$ and
	the steady state attained in six simulations run with different time
	steps $\Delta t$. Fig.~\ref{fig:stabilitypics2} shows the evolution
	of the associated $\dot{L}_1$ and the curves of the accomplished
	internal iterations at each time step. This group of simulations was
	run with GDBC.

	Since the semi-implicit scheme is unconditionally stable for any
	time step, the CPU time to reaching the desired numerical solutions
	can be be optimized without major restrictions to selected time
	step, as shown by Table~\ref{tab:comptimespent}. A conservative
	choice for all simulations would be $\Delta t = 0.1$, which presents
	the same $\dot{L}_1$ decay curve as for $\Delta t = 0.01$. The choice for
	the remainder of this work was $\Delta t = 0.5$ for the SH3
	simulations and $\Delta t = 0.1$ for the SH35, so that lower
	computational time is required without losing physically consistent
	transient results given the numerical scheme stability. 

	\begin{table}[H]
		\centering
		\caption{\label{tab:comptimespent}Computational time spent on
		each of the simulations presented in~\ref{fig:stabilitypics1}.
		simulations.}
		\vspace*{10pt}
		\centering
		\setlength\tabcolsep{4.0 pt}
		\renewcommand{\arraystretch}{1.0}
		\begin{tabular}{@{}lcccc@{}}
			\toprule [{1.5pt}]
			\textbf{Time}  & \textbf{Minimum} & \textbf{Maximum}
			& \textbf{Steady} & \textbf{Computational}\\[-8pt]
			\textbf{step}  & \textbf{iterations} & \textbf{iterations}
			& \textbf{state} & \textbf{time spent}
			\\
			\midrule
			$\Delta t=0.01$    & 1 & 23 & $t=3062$ & 234.4     \\[0.2mm]
    		$\Delta t=0.1$     & 2 & 15 & $t=3066$ & 6.6  \\[0.2mm]
    		$\Delta t=0.5$     & 4 & 18 & $t=3142$ & 1.7  \\[0.2mm]
    		$\Delta t=1.0$     & 5 & 22 & $t=3412$ & 1.4  \\[0.2mm]
    		$\Delta t=2.0$     & 7 & 23 & $t=4522$ & 1.2  \\[0.2mm]
    		$\Delta t=10.0$    & 7 & 16 & $t=30230$ & 1				
			\\
			\bottomrule [{1.5pt}]
		\end{tabular}
	\end{table}

\subsection{Convergence analysis}
	The adopted method to verify the order of accuracy of the code is the method
	of manufactured solutions (MMS). It provides a  convenient way of verifying
	the implementation of nonlinear numerical algorithms by using a manufactured
	(artificial) solution for such
	purpose~\cite{roache2002code,roy2005review,Vitral-2018}. In
	terms of the proposed problem,  we take all the members of the SH
	equation~\eqref{SH} and consider the following differential equation:

\begin{equation}
	G\left(\psi\right)\;=\;0,
	\label{eq:PDE}
\end{equation}
where $\psi$ is the order parameter function  that satisfies Eq.~\eqref{eq:PDE}
and therefore is the PDE solution. The MMS consists of adopting an arbitrary
function to be  the manufactured solution, $\psi_m(\mathbf{x},t)$, and since
this function is not likely to solve the PDE, a source term is expected,
$s_m$. This term can be seen as an additional forcing function, leading to a
modified operator with this new source:

\begin{equation}
	\bar{G}\left(\psi\right)\equiv G\left(\psi\right)-s_m.
	\label{eq:newPDE}
\end{equation}

\noindent For the previous equation, $\bar{G}\left(\psi_m\right)=0$ and
$\bar{G}\left(\psi\right)=-s_m$. Following this new approach to the problem,
we find an approximate numerical solution, $\psi_k$, for the discretized
problem so that $\bar{G}\left(\psi_k\right)=0$ or $G\left(\psi_k\right)=s_m$.
This source term is a minimal intrusion to the code’s formulation. The chosen
function is periodic with a wavenumber $q_1$ and is defined as:

\begin{equation}
	\psi_m(\mathbf{x},t)\;=\;\psi_0+\psi_{xy}\cos[q_1 (x+y)]e^{at},
	\label{eq:manufacturedsolution}
\end{equation}

\noindent where all parameters employed in the manufactured solution  and in the
differential equation are present on Table~\ref{tab:parametersMMS}.

\begin{figure}[htb!]
	\centering
	\includegraphics[width=0.9\textwidth]{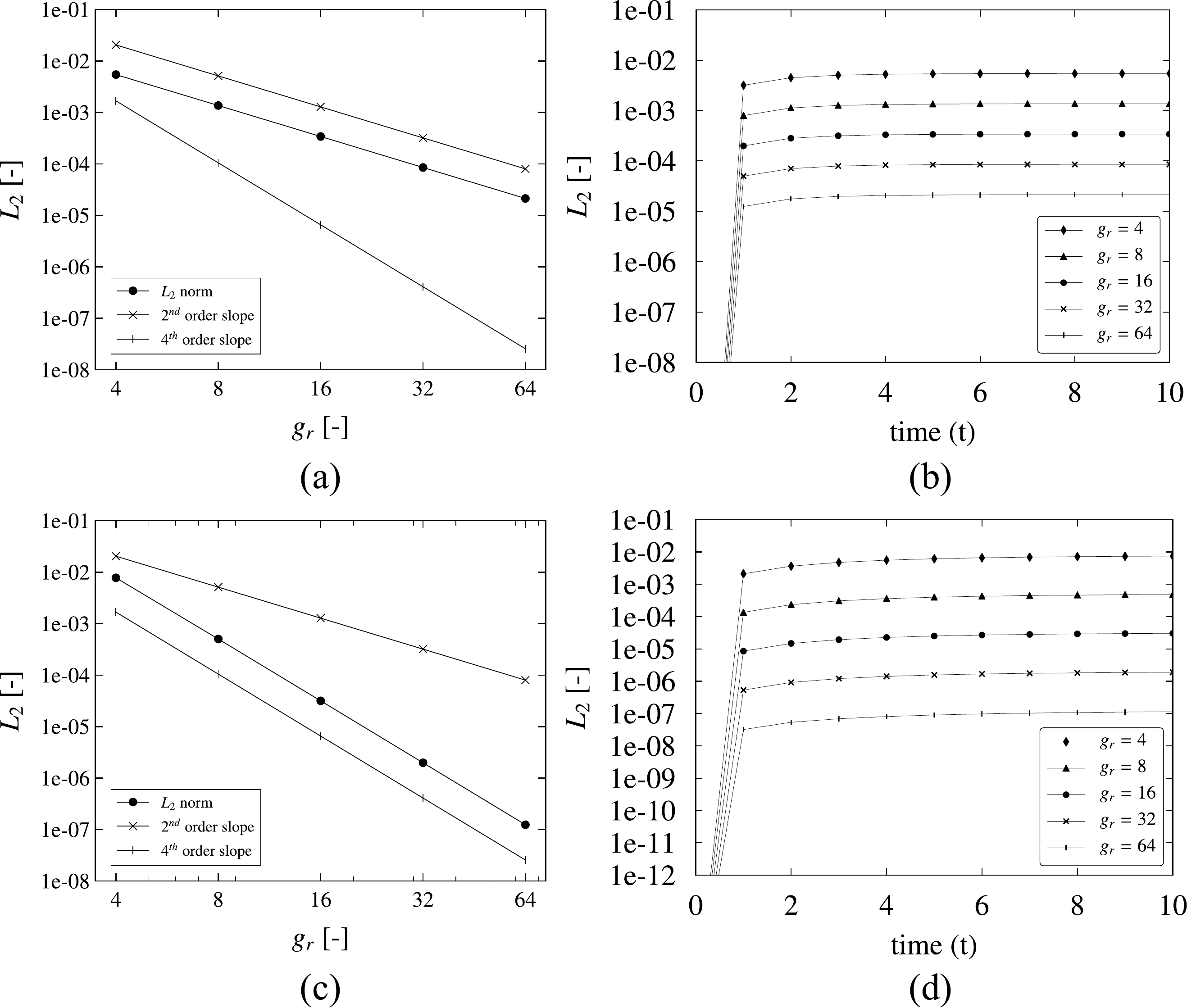}
	\caption{\label{fig:L2norm}The observed order of accuracy  of
	the code using Eq.~\eqref{eq:order} for both cases shown in
	Table~\ref{tab:parameterscode}. In first case, the $L_2$ norm is parallel to
	the (a) second-order slope and in the second case it is parallel to the (c)
	fourth-order slope. The resulting curves for the $L_2$ norm evolution in time
	are shown for the (b) first and (d) second case. Both $L_2$ norm evolutions
	start from values close to the machine precision ($10^{-16}$), since we start
	the convergence analysis simulations from the exact solution. In the first
	case the formal order of accuracy is achieved,  $p\approx 2.0$, while in the
	second case $p^{*}\approx 4.0$ due to  discretization error canceling. }
\end{figure}

\begin{table}[h]
	\centering
	\caption{\label{tab:parametersMMS}
	Parameters assumed for the manufactured solution adopted in the convergence
	analysis.
	}
	\vspace*{10pt}
	% \ars{1.2}
	\begin{tabular}{@{}c c c c@{}}
		\toprule [{1.5pt}]
		\textbf{MS Parameters} & \textbf{Value} & 
		\textbf{SH Parameters} & \textbf{Value}
		\\
		\midrule
		$q_1$ & $0.25\sqrt{2}q_0$ & $\epsilon$ & -0.1\\
		$q_{1}^{*}$ & $\sqrt{2}q_0$ & $q_0$ & 1.0\\
		$\psi_0$ & 0.0 & $\alpha$ & 1.0\\
		$\psi_{xy}$ & $\sqrt{|\epsilon|}$ & $\beta$ & -1.0\\
		$a$ & 0.0 & $\gamma$ & 0.0\\
		\bottomrule [{1.5pt}]
	\end{tabular}
\end{table}
The global discretization error was examined by the $L_2$ norm, defined as
follows:
\begin{equation}
	L_2\;=\;\left(
	\frac{\sum\limits_{i=1}^{n_x}\sum\limits_{j=1}^{n_y}\mid (\psi_k)_{i,j}^n-(\psi_m)_{i,j}^n\mid^2}
	{\sum\limits_{i=1}^{n_x}\sum\limits_{j=1}^{n_y}\mid (\psi_m)_{i,j}^n\mid^2} \right)^{1/2}.
	\label{eq:L2}
\end{equation}
	In the previous section, a second-order scheme was presented and
	therefore the formal order of accuracy is two. The observed order of
	accuracy of the code can be acquired from the global discretization
	error for meshes with different grid spacing, and can described by
	the following relation:
\begin{equation}
	p\;=\;\dfrac{\ln\left({L_2^B}/{L_2^A}\right)}{\ln(r)},
	\label{eq:order}
\end{equation}
	where $L_2^A$ and $L_2^B$ are the $L_2$ norm for meshes A and B respectively,
	and $r$ is the ratio of the grid resolution $g_r$, given in number of mesh
	nodes per critical wavelength of mesh B to A. The meshes employed for the
	present tests, the number of nodes, and grid resolutions are shown in
	Table~\ref{tab:parameterscode}, and the $L_2$ curves for the numerical
	experiments can be seen in Fig.~\ref{fig:L2norm}. 

\begin{table}[h]
	\centering
	\caption{\label{tab:parameterscode}
	Meshes employed for the code verification by MMS. The chosen domain for the
	test has $4\times4$ critical wavelengths. The observed  order of accuracy of
	the code in Fig.~\ref{fig:L2norm} can be  calculated using
	Eq.~\eqref{eq:order} for two cases: $q_1=0.25\sqrt{2}q_0$ and
	$q_{1}^{*}=\sqrt{2}q_0$. In the first case the formal order of accuracy is
	achieved,  $p\approx 2.0$, while in the second case $p^{*}\approx 4.0$ due to 
	discretization error canceling.
	}
	\vspace*{10pt}
	% \ars{1.2}
	\begin{tabular}{@{}l c c c c @{}}
		\toprule [{1.5pt}]
		\textbf{Mesh} & \textbf{Number} & \textbf{Grid} 
					  & \multicolumn{2}{c}{\textbf{Convergence rate} ($p$)}\\
		              & \textbf{of nodes} & \textbf{resolution} ($g_r$) & 
		              $\left(q_1=0.25\sqrt{2}q_0\right)$ & 
		              $\left(q_{1}^{*}=\sqrt{2}q_0\right)$
		\\
		\midrule
		Mesh A &  $16\times 16$     & 4         & - 			
		& - 		  \\[0.2mm] 
		Mesh B &  $32\times 32$     & 8         & $1.99514004$ 
		& $3.95294357$ \\[0.2mm] 
		Mesh C &  $64\times 64$     & 16        & $1.99862314$ 
		& $3.98876043$ \\[0.2mm] 
		Mesh D &  $128\times 128$   & 32        & $1.99964767$ 
		& $3.99722721$ \\[0.2mm] 
		Mesh E &  $256\times 256$   & 64        & $1.99991139$
		& $3.99916624$ \\[0.2mm] 
		\bottomrule [{1.5pt}]
	\end{tabular}
\end{table}

\clearpage

	A truncation error canceling could be seen when the choice for the
	manufactured solution wavenumber was $q_1=\sqrt{2}q_0$. The latter modifies the
	expected convergence rate up to fourth-order, which is consistent with the
	analytical development of the manufactured solution in Eq.~\eqref{eq:newPDE}.
	To verify the general case, we construct a manufactured solution with $q_1
	\neq \sqrt{2}q_0$. Following this approach, a second-order convergence rate is
	recovered as theoretically expected. 

	The numerical scheme has been proven to be consistent, unconditionally stable
	and truly second-order accurate in space, in the general case. The chosen
	grid resolution for the numerical experiments, presented in the next section,
	was $g_r=16$, which has a good trade-off between resolution and computational
	cost in order to represent periodic solutions.

\subsection{Lyapunov functional decay}

\begin{figure}[htb!]
	\centering
	\includegraphics[width=0.9\textwidth]{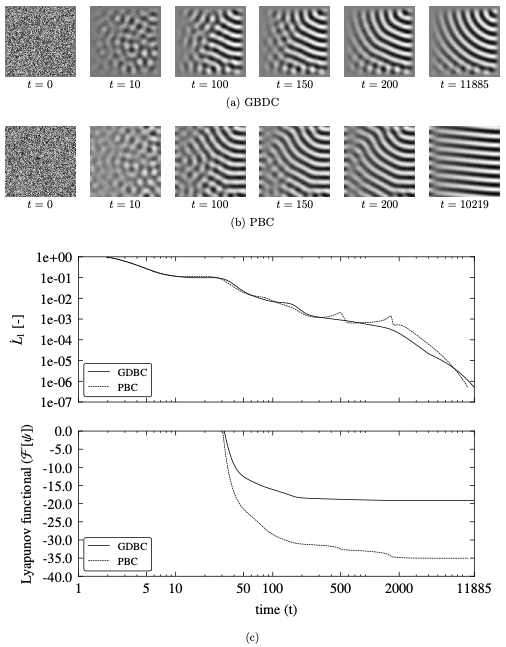}
	\caption{\label{fig:lyaresults1}The results of two simulations run to
	verify the correctness of the implementation of the Lyapunov
	functional for the SH3 model. First and second rows: SH3
	pattern evolution for a ramped system forced with
	$0.0\leq\epsilon\leq 0.2$ GDBC (a) and PBC (b), respectively, until the indicated steady state.. Both
	simulations started with the same pseudo-random initial condition.
	$\dot{L}_1$ and the Lyapunov potential evolutions are showed below.}
\end{figure}

\begin{figure}[H]
	\centering
	\includegraphics[width=0.9\textwidth]{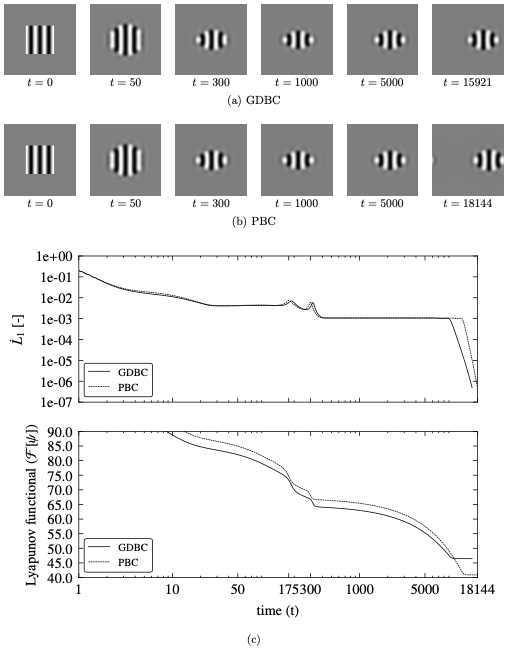}
	\caption{\label{fig:lyaresults2}The results of two simulations run to
	verify the correctness of the implementation of the Lyapunov
	functional for the SH35 model. First and second rows: SH35 pattern
	evolution for a ramped system forced with $-1.5\leq\epsilon\leq
	-1.4$, GDBC (a) and PBC (b), respectively, until the indicated steady state.. Both simulations started
	from the same pre-existing structure of rolls perpendicular to the
	gradient of the control parameter. $\dot{L}_1$ and the Lyapunov potential
	evolutions are showed below.}
\end{figure}

\begin{figure}[H]
	\centering
	\includegraphics[width=0.9\textwidth]{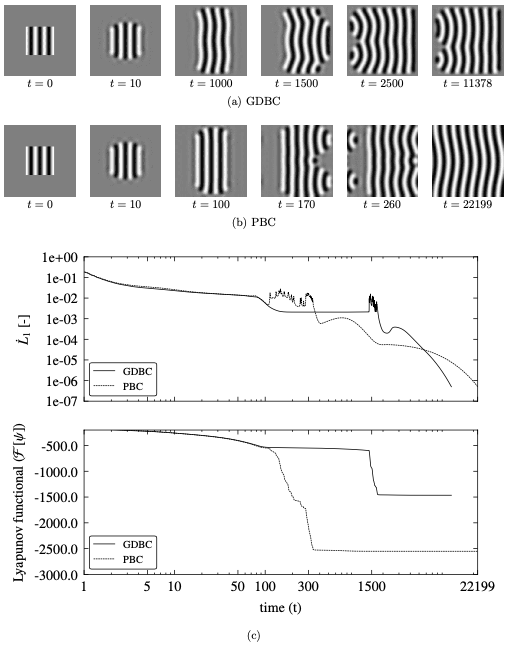}
	\caption{\label{fig:lyaresults3}The results of two simulations run to
	verify the correctness of the implementation of the Lyapunov
	functional for the SH35 model. First and second rows: SH35 pattern
	evolution for a ramped system forced with $-1.4\leq\epsilon\leq
	-1.2$, GDBC (a) and PBC (b), respectively, until the indicated steady state.. Both simulations started
	from the same pre-existing structure of rolls perpendicular to the
	gradient of the control parameter. $\dot{L}_1$ and the Lyapunov potential
	evolutions are showed below.}
\end{figure}

	The decay of the ``free energy'' (Lyapunov) functional is monitored so that we
	can confirm it is always monotonically decreasing until a minimum value
	is reached in the steady state. In order to take into account the assumed
	boundary conditions adopted in this work,  we expand the bulk integrals to
	show that boundary integrals vanish, leading to a new expression for the free
	energy functional, which is discretized. Using integration by
	parts and the 2D Gauss theorem (first Green identity), we have:
	\begin{equation}
		\int_\Omega d\mathbf{x} \left(\psi\nabla^2\psi \right)=
		\oint_{\partial\Omega} \psi \dfrac{\partial\psi}{\partial n}dl
		- \int_\Omega d\mathbf{x} |\nabla\psi|^2 \,,
		\label{eq:boundary_LF1}
	\end{equation}
	where the first integral from the RHS vanishes for the assumed GDBC ($\psi=\partial\psi/\partial n=0 \;\text{on}\; \partial\Omega$) and PBC (since ``there are no boundaries''). Now Eq.~\eqref{eq:LF1} can be rewritten as:
	\begin{equation}
		\mathcal{F}[\psi]=
		\int_\Omega {d\mathbf{x}}\dfrac{1}{2}
		\left\{
			-{\epsilon(\mathbf{x})}\psi^2+{\alpha}
			\left[
				q_{0}^4\psi^2-2q_{0}^{2}(\nabla\psi)^2+
				(\nabla^2\psi)^2
			\right]
			-\dfrac{\beta}{2}\psi^4+\dfrac{\gamma}{3}\psi^6
		\right\}.
		\label{eq:LF2}
	\end{equation}
	The Lyapunov functional associated to the SH equation is implemented
	through the discrete formula derived by Christov \&
	Pontes~(2001)\cite{C.I-2002} for the cubic version, and now extended
	for the quintic one. The formula presents a ${\cal O}\left(\Delta
	t^2+\Delta x^2+ \Delta y^2\right)$ approximation of the functional
	given by Eq.~\eqref{eq:dec_LF}. As pointed by those authors, the
	monotonic decay of the finite differences version is enforced,
	provided that the internal iterations converge.{Observe that the rate of change of the functional for this choice of boundary conditions (GDBC and PBC) is given by
	\begin{equation}
		\dfrac{d\mathcal{F}}{dt} = \int_\Omega d\mathbf{x}\,
		\mu\dfrac{\partial\psi}{\partial t} =
		- \int_\Omega d\mathbf{x} \, |\mu|^2 \leqslant 0 \,,
	\end{equation}
	which in time discrete form reads
	\begin{align}
	 \frac{{\cal F}^{n+1} - {\cal F}^n}{\Delta t}
		= \int_\Omega d\mathbf{x} \, \mu^{n+1/2}
		\left({\psi^{n+1}- \psi^n\over \Delta t}\right) 
		+ E_r \,,
		\label{eq:Lyapu_diff}
	\end{align}
	such that the RHS is guaranteed to be always negative. To show the consistency of the scheme the error $E_r\propto\mathcal{O}(\Delta t^\eta)$, for $\eta> 0$.
	Hence, the gradient descent nature of the dynamical equation ensures that any consistent discretization must
	preseve the correspondent energy disspation property.
	
	In other to show the latter, we input the RHS of the target scheme
	given by $\mu^{n+1/2}$ in Eq.~\eqref{eq:numericalscheme} into Eq.~\eqref{eq:Lyapu_diff}:
	\begin{align}
	 E_r = \dfrac{1}{\Delta t} 
	 \left[{{\cal F}^{n+1} - {\cal F}^n}
		- \int_\Omega d\mathbf{x} \, \mu^{n+1/2}
		\left({\psi^{n+1}- \psi^n}\right)\right] \,,
		\label{eq:Lyapu_diff}
	\end{align}
	
    \begin{align}
        \Delta t  E_r = &{\cal F}^{n+1} - {\cal F}^n
        - \int_\Omega d\mathbf{x} \, 
		\dfrac{1}{2}
		\left[
			\epsilon(\mathbf{x})
			-\alpha\left(q_0^2+\nabla^2\right)^2
			+\beta\dfrac{\left(\psi^{n+1}\right)^2
			+
			\left(\psi^n\right)^2}{2}
			-
			\gamma\dfrac{\left(\psi^{n+1}\right)^4
			+
			\left(\psi^n\right)^4}{2}
		\right]\times
		\nonumber\\&
		\left[({\psi^{n+1})^2 - (\psi^n)^2}\right] \nonumber\\
		=\,& - \int_\Omega d\mathbf{x} \left[\, 
		\frac{\gamma \left(\psi^{n+1}\right)^{6}}{12} - 
		\frac{\gamma \left(\psi^{n+1}\right)^{4} \left(\psi^{n}\right)^{2}}{4} 
		+ \frac{\gamma \left(\psi^{n+1}\right)^{2} \left(\psi^{n}\right)^{4}}{4} 
		- \frac{\gamma \left(\psi^{n}\right)^{6}}{12} \right.\nonumber\\&\left.
		- \alpha q_{0}^{2} \psi^{n+1}
		\nabla^2 \psi^{n} + a q_{0}^{2} \psi^{n} \nabla^2 \psi^{n+1} - \frac{\alpha}{2}{\psi^{n+1} \nabla^4 \psi^{n}} + \frac{\alpha}{2}{\psi^{n} \nabla^4 \psi^{n+1}} \right]
	\end{align}
	
	Considering the GDBC and PBC, the divergent terms vanish (first Green identity) yielding the following error:
	\begin{align}
			E _r=\, \Delta t^2 \int_\Omega d\mathbf{x} \left[
		\gamma\frac{\left(\psi^{n+1} + \psi^{n}\right)^3}{12} \left(\mu^{n+1/2}\right)^3 \right]
	\end{align}
	
	\noindent Therefore, we have shown that $E _r= O(\Delta t^\eta)$, with $\eta=2$, and the discrete scheme is consistent, with second order accuracy.
	
	Following Eq.~\eqref{eq:LF2}, the discrete functional at each time step is
	given by:}
	\begin{align}
		{\cal F}^n =&
		\sum\limits_{i=1}^{n_x}\sum\limits_{j=1}^{n_y}
		\left[-{\epsilon\over 2}\left(\psi^{n}_{i,j}\right)^2
		-
		{\beta\over 4}
		\left(\psi^{n}_{i,j}\right)^4
		+
		{\gamma\over 6}
		\left(\psi^{n}_{i,j}\right)^6
		+{\alpha q_0^4\over 2} \left(\psi^{n}_{i,j}\right)^2 \right]
		\nonumber
		\\
		-\frac{2\alpha q_0^2}{4}
		\sum\limits_{i=1}^{n_x}\sum\limits_{j=1}^{n_y}
		&
		\left[
			{\psi^{n}_{i+1,j}-\psi^{n}_{i,j}\over \Delta x}
		\right]^2
		\!\!\!+\!
		\left[
			{\psi^{n}_{i,j}-\psi^{n}_{i-1,j}\over \Delta x}
		\right]^2
		\!\!\!+\!
		\left[
			{\psi^{n}_{i,j+1}-\psi^{n}_{i,j}\over \Delta y}
		\right]^2
		\!\!\!+\!
		\left[ 
			{\psi^{n}_{i,j}-\psi^{n}_{i,j-1}\over \Delta y}
		\right]^2
		\nonumber
		\\
		+\frac{\alpha}{2}
		&
		\sum\limits_{i=1}^{n_x}\sum\limits_{j=1}^{n_y}
		\left[
			{\psi^{n}_{i+1,j}-2\psi^{n}_{i,j}+\psi^{n}_{i-1,j}
			\over \Delta x^2}
			+
		 	{\psi^{n}_{i,j+1}-2\psi^{n}_{i,j}+\psi^{n}_{i,j-1}
			\over \Delta y^2}
		\right]^2\,.
		\label{eq:discfunc}
	\end{align}
	{
	In order to verify that the numerical scheme is correctly implemented,
	we monitored the evolution of the Lyapunov functional (Eq.~\eqref{eq:discfunc}) in all the performed simulations, and verified that the Lyapunov potential is always decreasing.
	As an example, we show}
	two simulations with the cubic SH (with ramped $\epsilon$) and four with the
	quintic one (both uniformly and ramped $\epsilon$), and the
	evolution of the {Lyapunov functional (Eq.~\eqref{eq:discfunc})}, of $\dot{L}_1$ and of the patterns observed
	through the order parameter $\psi(\mathbf{x},t)$. The patterns evolution and
	the associated $\dot{L}_1$ and Lyapunov potentials are presented in
	Figs.~\ref{fig:lyaresults1}, \ref{fig:lyaresults2}, and
	\ref{fig:lyaresults3}. The simulations were run with ramps of the control
	parameter $\epsilon(\mathbf{x})$ along the $x$-direction for both GDBC and
	PBC. The two simulations with the SH3 model started from the same
	pseudo-random initial condition. The four simulations with the SH35 model
	started from a squared localized patch in the center of the cell. This
	pre-existing structure of stripes parallel to the $y$-direction were
	constructed by:
	\begin{equation}
		\psi(\mathbf{x},0)=A_0\cos(q_0 x),
	\end{equation}

	\noindent
	where $A_0=\left\{(\beta/(2\gamma))[1+\sqrt{1+\epsilon\gamma/
	\beta^2}]\right\}^{1/2}$, which is compatible with the expected
	field $\psi$ amplitude for a stable spatially homogeneous states,
	for the SH35~\cite{sakaguchi1996}. The results  of these six
	simulations confirm the correct implementation of the numerical scheme for both equations, with a monotonic decay of the Lyapunov functional in all cases.

\begin{figure}[H]
	\centering
	\includegraphics[width=0.9\textwidth]{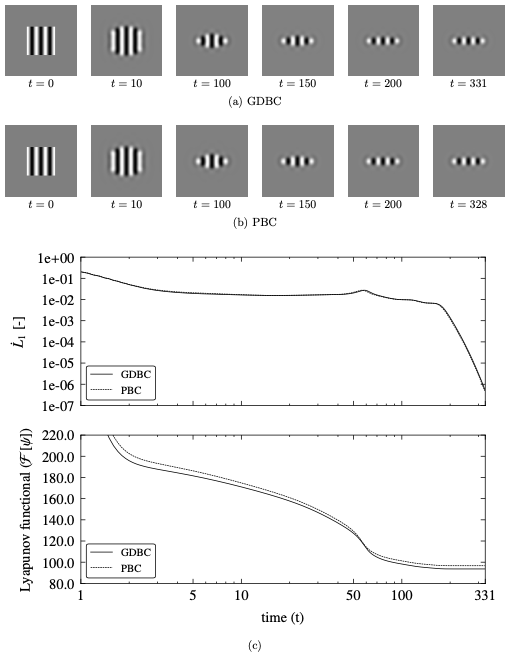}
	\caption{\label{fig:lyaresults2a}The results of two simulations run
	to verify the correctness of the implementation of the Lyapunov
	functional for the SH35 model. First and second rows: SH35 pattern
	evolution for a uniformly forced system with $\epsilon=-1.5$, GDBC
	(a) and PBC (b), respectively, until the indicated steady state.. Both simulations started from the
	same pre-existing structure of rolls perpendicular to the gradient
	of the control parameter. $\dot{L}_1$ and the Lyapunov potential
	evolutions are showed below.}
\end{figure}

\section{\label{sec:numerical experiments}Numerical experiments}
	The numerical scheme described in Sec.~\ref{sec:mathmodel} is now used
	to solve the SH3 and the SH35 equations in square domains with rigid
	and periodic boundary conditions, and with nonuniform forcings. The adopted
	parameter values are given in Tables ~\ref{tab:param}, ~\ref{tab:parameters1} and
	\ref{tab:parameters2}. All simulations were done in a
	satisfactory grid resolution of the spatial grid of $16$ points per
	critical wavelength ($g_r=16$).
\subsection{\label{sec:numerical SH3}Numerical experiments with the SH3
equation}
\begin{figure}[H]
	\centering
	\includegraphics[width=0.9\textwidth]{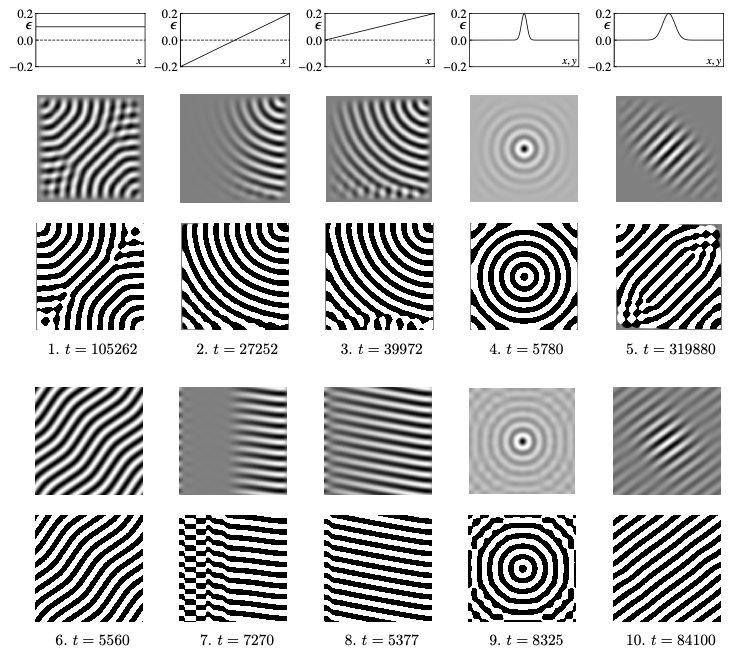}
	\caption{\label{fig:summary1} Zoology of ten patterns obtained by 
		numerical integration of the SH3 equation with GDBC (rows 1 and 2) and PBC
		(rows 3 and 4). All patterns shown correspond to the steady attained when
		$\dot{L}_1\leq 5.0\times10^{-7}$ (Eq.~\eqref{eq:L1}). All simulations started
		from the same pseudo-random initial condition. The applied forcing is shown
		in the first row. First column: uniform forcing; Second and third columns:
		ramps of the control parameter $\epsilon$; Fifth and sixth columns: Two
		Gaussian forcings, according to Eq.~\eqref{distgauss}. The first and third rows
		correspond to GDBC and PBC, respectively. Second and fourth rows contains
		the same patterns shown in rows one and three, respectively, with an
		enhanced contrast. Patterns configurations 2 and 3 tend to approach perpendicularly to
		the supercritical boundaries and parallel to subcritical ones. The pattern
		configuration 4 presents a target structure compatible with the narrow Gaussian
		profile of the forcing. The target collapses in the pattern configuration 5, and
		evolves to a structure of rolls parallel to one of diagonals, and not to the
		side walls. This effect occurs as a consequence of the modulation, which if
		harder along the direction of the pattern wavevector. Similar phenomena
		occurs with patterns configurations 9 and 10.} 
\end{figure}

	In the case of the SH3 equation we present the result of ten
	simulations (denoted as configurations $1$ to $10$), five of them
	with GDBC and five with PBC. The resulting steady states for each simulation
	are summarized in Fig.~\ref{fig:summary1}. All simulations started
	with the same pseudo-random initial condition. For each type of
	boundary condition we ran a configuration with uniform forcing to
	reproduce existing results, two configurations with ramps of the
	control parameter $\epsilon$, as detailed in
	Table~\ref{tab:parameters1}, and two ones with different Gaussian
	distributions of the control parameter. In ramped systems, the
	parameter is assumed as uniform along the $y$-direction. The Gaussian
	distribution of $\epsilon$ introduces symmetric gradients in the
	radial direction.

	The steady state solutions obtained are shown in
	Fig.~\ref{fig:summary1}. The first line in this figure shows the
	distribution of the control parameter $\epsilon$ along the
	$x$-direction of the domain. Second and fourth lines present the
	steady state solutions with rigid and with periodic boundary
	conditions, respectively. Third an fifth lines present the same
	results shown in lines two and four, but this time with frames
	constructed with enhanced contrast to make visible the subcritical
	structure developed in regions where the control parameter $\epsilon$
	is negative. Each configuration run is identified with an
	assigned number.

\begin{table*}[h]
	\centering
	\caption{\label{tab:parameters1}Parameters assumed for the
	numerical experiments with the SH3 equation.}
	\vspace*{10pt}
	\begin{tabular}{@{}c c c c c c l @{}}
		\toprule [{1.5pt}]
		\textbf{Parameter} && \textbf{Formul\ae} && \textbf{Value} 
		&& \textbf{Description}
		\\
		\midrule
		$q_0$ && -   && 1.0        && Critical wavenumber
		\\[0.2mm]
		$\lambda_0$  && $2\pi/q_0$ && $2\pi$ && Critical wavelength
		\\[0.2mm]
		$w_x$, $w_y$ &&  - && 10   && Wavelengths per domain length
		\\[0.2mm]
		$g_r$  && - && 16 && Grid resolution (nodes per wavelength)
		\\[0.2mm]
		$n_x$, $n_y$ &&  $w_x\times g_{r}$,$w_y\times g_{r}$ && 160
		&& Nodes per mesh side ($n$)
		\\[0.2mm]
		$N$    &&  $n_x\times n_y$ && $160\times 160$
		&& Total number of mesh nodes
		\\[0.2mm]
		$L_x$, $L_y$ &&  $w_x \lambda_0$, $w_y \lambda_0$ &&
		$\approx 62.832$  && Square domain length ($L$)
		\\[0.2mm]
		$\Delta x$,$\Delta y$ &&   $L/(n-2)$ && $\approx 0.3977$ &&
		Space step (GDBC)
		\\[0.2mm]
		$\Delta x$,$\Delta y$ &&   $L/n$ && $\approx 0.3927$     &&
		Space step (PBC)
		\\[0.2mm]
		$\Delta t$  &&  - && 0.5 && Time step for SH3
		\\[0.2mm]
		$A$  && - && 0.2 && Gaussian maximum value (peak)
		\\[0.2mm]
		$R_1$  && $N$ && - && For configurations 4 and 9
		\\[0.2mm]
		$R_2$  && $0.2N$ && - && For configurations 5 and 10
		\\[0.2mm]
		$x_0$,$y_0$  && $L_x/2$,$L_y/2$ && - && Gaussian center
		\\[0.2mm]
		\bottomrule [{1.5pt}]
	\end{tabular}
\end{table*}

	Configuration 1 shows a pattern emerging in a uniformly forced system with
	GDBC, in qualitative agreement with previous
	results~\cite{Greenside1984,Manneville1990,Cross1993,pontes2008}. This case
	was run for further verification of our numerical code. A pattern with
	unavoidable defects is developed, in order to match the strong requirement of
	rolls approaching perpendicularly the side walls. A structure with lower
	density of defects appears in configuration 6, where the same forcing of
	Frames 1 is assumed, but now, with periodic boundary conditions. The pattern
	develops a zig-zag instability.  

	Configuration 2 reproduces the result of a pattern developed in presence of a
	ramped control parameter $\epsilon$, with a subcritical
	region~\cite{C.I-1997,C.I-2002,pontes2008}. A structure with smaller amount
	of defects emerges, thanks to the fact that the weak structure of rolls
	parallel to the wall appears in the subcritical region. A similar situation
	occurs in configuration 3, where the system is forced with a ramp of the
	control parameter $\epsilon$ , however taking non negative values. In this
	case a weak structure of rolls perpendicular to the lower $(y=0)$ sidewall
	is visible.

	Configurations 7 and 8 present the results for systems with same
	forcing of configurations 2 and 3, for the case of periodic boundary
	conditions. A tendency to develop a structure of rolls parallel to
	the gradient of the control parameter $\epsilon$ clearly appears, as
	well as the existence of a Benjamin-Feir instability close to the
	left wall, induced by the the supercritical region close to the
	right wall.

	Configurations 4, 5, 9 and 10  present numerical results of simulations
	performed with a Gaussian radial distribution of the control parameter
	$\epsilon$, in the form:

	\begin{equation}
		\epsilon(\mathbf{x})=Ae^{-R((x-x_0)^2+(y-y_0)^2)} \;,
		\label{distgauss}
	\end{equation}

	\noindent where the parameters are specified in Table~\ref{tab:parameters1}. The
	parameter $R$ is assumed to be $R_1$ in configurations 4 and 9; and $R_2$ in
	configurations 5 and 10. Worth noticing that these Gaussian distributions
	introduce a gradient of $\epsilon$ in all directions with the maximum of
	$\epsilon$ at the center of the domain.

	Configurations 4 and 5 were run with rigid boundary conditions and
	configurations 9 and 10, with periodic conditions. A sharp Gaussian,
	rapidly decreasing from the maximum value across a short distance,
	leads to the onset of a target pattern with both boundary
	conditions. The target pattern collapses in presence of a wider Gaussian
	distribution of the control parameter $\epsilon$, leading to a
	structure of rolls. The structure wavevector takes the direction of
	one of the domain diagonals, again, since the most difficult
	direction for modulations of the structure is the one associated to
	the wavevector, whereas the easiest one is the direction
	perpendicular to the wavevector.

\begin{figure}
	\centering
	\includegraphics[width=0.9\textwidth]{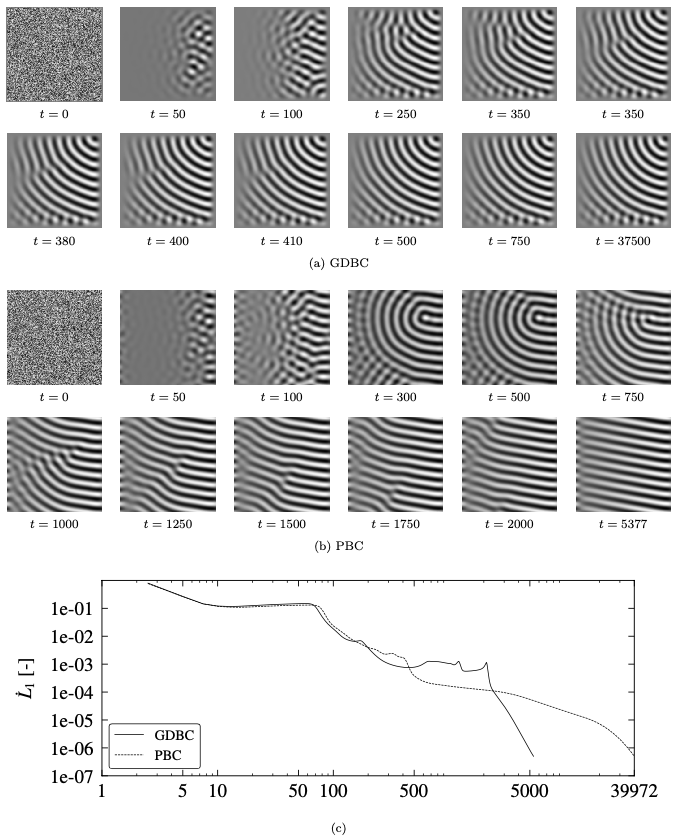}
	\caption{\label{fig:evo2}First and second lines: pattern evolution
	for the configuration 3 of Fig.~\ref{fig:summary1}, with GDBC (a),
	until the steady state. Third and fourth lines: pattern evolution
	for the configuration 8 of Fig.~\ref{fig:summary1}, with PBC (b),
	until the steady state. In both cases, the system is forced with a
	ramp of the control parameter $\epsilon$ along the $x$ direction,
	with $-0.2\leq\epsilon\leq 0.2$. The associated $\dot{L}_1\times t$ curves
	are also shown (c).}
\end{figure}

\begin{figure}[H]
	\centering
	\includegraphics[width=0.9\textwidth]{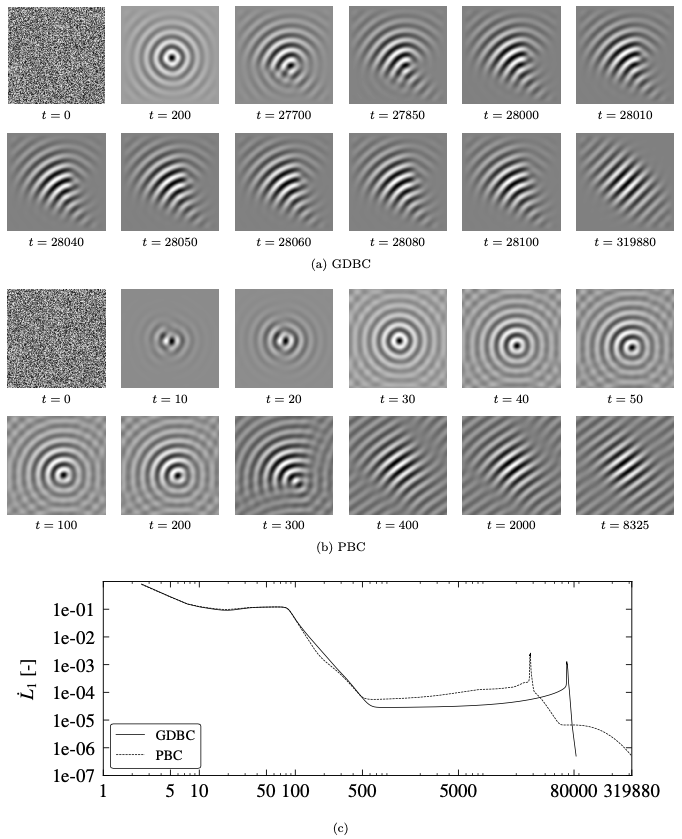}
	\caption{\label{fig:evo4}First and second lines: pattern evolution
	for the configuration 5 of Fig.~\ref{fig:summary1}, with GDBC (a),
	until the steady state. Third and fourth lines: pattern evolution
	for the configuration 10 of Fig.~\ref{fig:summary1}, with PBC (b),
	until the steady state. In both cases, the system is forced with a
	Gaussian distribution  of the control parameter $\epsilon$. The
	associated $\dot{L}_1\times t$ curves are also shown (c).}
\end{figure}

	The time evolution of configurations 3 (GDBC) and 8 (PBC), from the
	pseudo-random initial condition to the steady state is shown in
	Fig.~\ref{fig:evo2}. The evolution of configurations 5 (GDBC) and
	10 (PBC) are shown in Fig.~\ref{fig:evo4}. All curves reveal the
	expected result, with $\dot{L}_1$ rapidly decreasing as a pattern emerges
	and the amplitude of the structure essentially saturates. This stage
	is followed by a much slower phase dynamics, ending with an
	exponential decay towards the steady state. In general, $\dot{L}_1$ peaks
	are associated to the removal of defects.
	We observe that boundary conditions have little effect on the
	resulting curves.

\subsection{\label{sec:numerical SH35}Numerical experiments with the SH35
equation}
	The quintic version of the Swift-Hohenberg equation with destabilizing cubic
	term allows for the existence of stable localized solutions for an interval
	of subcritical values of~$\epsilon$, i.e. coexistence of the solution
	$\psi=0$ and a modulated one, shown by ~\cite{sakaguchi1996}.  The parameter
	values adopted are given in Table~\ref{tab:parameters2}. In this sense, we run
	six experiments with the SH35 equation: two with uniform forcing
	$(\epsilon=-1.5)$, GDBC and PBC, respectively, two with a ramp
	$-1.5\leq\epsilon\leq -1.4$, GDBC and PBC, respectively, and two last ones
	with a ramp $-1.4\leq\epsilon\leq-1.2$, GDBC and PBC, respectively. The
	experiments with uniform forcing were conducted as code verification, to
	reproduce existing results. The time evolution of these two experiments are
	shown in Fig.~\ref{fig:lyaresults2a}. Both simulations started from the same
	pseudo-random initial condition.

	The four experiments with ramped systems started from the same
	pre-existing structure of rolls perpendicular to the gradient of the
	control parameter. These experiments were used to verify the
	implementation of the Lyapunov functional with the quintic equation
	SH35. The evolution of these simulations are shown in
	Fig.~\ref{fig:lyaresults2}, and \ref{fig:lyaresults3}. In the case
	of a ramp given by $-1.5\leq\epsilon\leq -1.4$ the pre-existing
	structure evolves towards a localized structure of rolls with the
	same orientation of the initial condition, with both boundary
	conditions. When the applied forcing takes the form
	$-1.4\leq\epsilon\leq-1.2$, lowering the energy associated with the
	non-trivial solution both with GDBC and PBC, a structure of rolls
	with the same orientation of the initial condition also prevails, but
	the final structure is localized with GDBC and occupies the entire
	domain, with PBC.

	\begin{table*}[h]
		\centering
		\caption{\label{tab:parameters2}Parameters assumed for the
		numerical experiments with the SH35 equation.}
		\vspace*{10pt}
		% \ars{1.2}
		\begin{tabular}{@{}c c c c c c l @{}}
			\toprule [{1.5pt}]
			\textbf{Parameter} && \textbf{Formul\ae} && \textbf{Value} 
			&& \textbf{Description}
			\\
			\midrule
			$q_0$ && -   && 1.0        && Critical wavenumber
			\\[0.2mm]
			$\lambda_0$  && $2\pi/q_0$ && $2\pi$ && Critical wavelength
			\\[0.2mm]
			$w_x$, $w_y$ &&  - && 8   && Wavelengths per domain length
			\\[0.2mm]
			$g_r$  && - && 16 && Grid resolution (nodes per wavelength)
			\\[0.2mm]
			$n_x$, $n_y$ &&  $w_x\times g_{r}$,$w_y\times g_{r}$ && 128
			&& Nodes per mesh side ($n$)
			\\[0.2mm]
			$N$    &&  $n_x\times n_y$ && $128\times 128$
			&& Total number of mesh nodes
			\\[0.2mm]
			$L_x$, $L_y$ &&  $w_x \lambda_0$, $w_y \lambda_0$ &&
			$\approx 50.266$  && Square domain length ($L$)
			\\[0.2mm]
			$\Delta x$,$\Delta y$ &&   $L/(n-2)$ && $\approx 0.3989$ &&
			Space step (GDBC)
			\\[0.2mm]
			$\Delta x$,$\Delta y$ &&   $L/n$ && $\approx 0.3927$     &&
			Space step (PBC)
			\\[0.2mm]
			$\Delta t$  &&  - && 0.1 && Time step for SH35
			\\[0.2mm]
			\bottomrule [{1.5pt}]
		\end{tabular}
	\end{table*}

	\begin{figure}[H]
		\centering
		\includegraphics[width=0.6\textwidth]{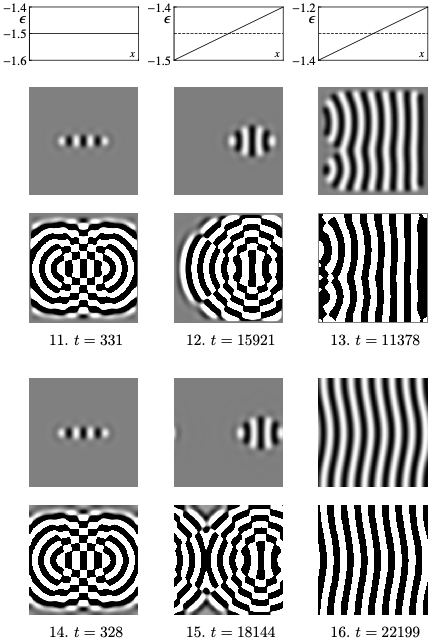}
		\caption{\label{fig:summary2} Zoology of six patterns obtained by 
			numerical integration of the SH35 equation with GDBC (rows 1
			and 2) and PBC (rows 3 and 4). All patterns shown correspond
			to the steady attained when $\dot{L}_1\leq 5.0\times10^{-7}$
			(Eq.~\eqref{eq:L1}) All simulations started from a
			pre-existing structure of stripes oriented along the $y$
			direction. The applied forcing is shown in the first row.
			First column: uniform forcing with $\epsilon=-1.5$. Second
			column: forcing with ramp of the control parameter in the
			form $-1.5\leq\epsilon\leq-1.4$. Third column: ramp in the form
			$-1.4\leq\epsilon\leq-1.2$. The first and third rows
			correspond to GDBC and PBC, respectively. Second and fourth
			rows contain the same patterns shown in rows one and three,
			respectively, with an enhanced contrast. In all cases, the
			pre-existing structure of stripes oriented along the $y$
			prevails at the steady state. However, configurations
			11, 12, 14 and 15 evolved to localized structures, whereas,
			in the case of configurations 13 and 16 the resulting
			pattern occupy the entire domain.} 

	\end{figure}

\subsection{\label{sec:discussion}Discussion}
	We extended a numerical scheme proposed by Christov \& Pontes~\cite{pontes2008} to investigate pattern formation modelled by the cubic SH equation in two dimensions. 

	Our work includes the quintic SH
	equation and PBC for both the cubic and the quintic SH, with a scheme that retains all characteristics of the original one:
	strict representation of the Lyapunov functional,
	unconditional stability, and second order representation of all
	derivatives. We additionally implement different nonuniform forcing field, such as ramps and Gaussian distributions.
	A detailed study of the effects of nonuniform forcings
	are addressed in a paper that builds on the present numerical scheme~\cite{coelho2021stripe}.

	Verification tests consisted in qualitatively reproducing existing
	results of SH3 and SH35 simulations with uniform
	forcings for both GDBC and PBC, where good agreement was observed. As an
	additional verification procedure, we also found good agreement
	between the results obtained from the SH3 forced with a spatial ramp of
	$\epsilon$, and those presented by Pontes \emph{et al.}
	(2008)\cite{pontes2008}. In that work, the authors adopted GDBC
	and a first order approximation in time. The results of the SH3 equation
	with GDBC and ramps of $\epsilon$ are presented in
	Figs.~\ref{fig:stabilitypics1}, \ref{fig:lyaresults1}a and in
	simulations configurations 1, 2 and 3 of Fig.~\ref{fig:summary2}.

	Rigid boundary conditions (GDBC) make patterns of supercritical
	stripes developed close to different boundaries to compete, due to
	the tendency of stripes to approach boundaries perpendicularly to
	them. That is, stripes cannot be perpendicular to all boundaries
	simultaneously. In addition, boundary effects compete with bulk
	ones. The overall result is a structures with high density of
	defects. A different situation occurs when PBC are imposed. In this
	case, a periodicity is imposed to the emerging pattern, and boundary
	effects are excluded. The result is, in general, a pattern with
	fewer defects than ones developed with the same forcing and GDBC.
	This is also the case when the system is forced with nonuniform
	distributions of the control parameter.

	An important configuration in nonuniformly forced systems consists
	in adopting GDBC, and a forcing where part of the domain is
	maintained at subcritical conditions. In this case a less known
	result points to the fact that the subcritical pattern of stripes,
	induced by the supercritical region, approach sidewalls parallel to
	them. These patterns are automatically perpendicular to adjacent
	walls. As a result, nonuniformly forced systems with a subcritical
	region tend to develop patterns with a lower density of defects than
	systems uniformly forced~\cite{pontes2008}. The presented
	simulations reproduce this effect.

	A particular feature observed in nonuniformly forced systems, 
	using the SH3 equation, PBC and spatial ramps of the control
	parameter, consists in the development of several examples of
	the Benjamin-Feir instability, which manifest itself in the
	form of ``alternating'' rolls, in a dislocation sense,
	promoted by a region where minimum and maximum values of
	$\epsilon$ are close.  
	
	The quintic SH35 equation with a destabilizing cubic term allows
	for the emergence of localized solutions in an interval of
	subcritical values of $\epsilon$. Six simulations were run
	with the quintic equation, assuming three different forcings
	and two boundary conditions for each one. All simulations 
	started from the same pseudo-random initial condition. The
	resulting steady states are presented in Fig.~\ref{fig:summary2}.
	The first group consisted of two simulations with uniform
	forcing $\epsilon=-1.5$. These simulations are denoted by
	configurations 11 and 14 in Fig.~\ref{fig:summary2}
	The second group was forced with a ramp along the $x$
	direction, given by $-1.5\leq\epsilon\leq -1.4$ (configurations
	12 and 15). The third group was run with $-1.4\leq\epsilon\leq -1.2$
	(configurations 13 and 16). In all configurations, the resulting
	patterns consisted of stripes perpendicular to the
	gradient of the control parameter, with localized patterns
	in the first and second groups, and the stripes occupying the
	entire domain in the third group. A mesh of $g_r=16$ points per critical wavelength was adopted in all
	simulations, which represents a good trade-off between spatial
	resolution and computational cost.

    The uniqueness of the numerical solutions of the SH equation has not been proved. We note that both the explicit terms, and the operators allocated to the implicit ones contain nonlinear terms. The scheme preserves both the intrinsic ability of nonlinear dynamics of generating new modes, and the sensitivity to the prescribed initial condition. Due to that, minor changes in the initial condition may change the final patterns. However, this property is restricted by geometric, bulk, boundary effects, reducing the possible steady states. Bulk effects include the level of forcing, gradients of the control parameter and imposed periodicity. For instance we can expect that the patterns developed in configurations 1 to 5 of Fig.~\ref{fig:summary1} could be mirror images with respect to the lower horizontal boundary, but not much more. The level of forcing is low and boundary effects prevail. The number of possible solutions is strongly reduced. At higher forcings small changes in the initial conditions may result in patterns with quantitatively different distribution of defects. The number of solutions is larger. In addition, some solutions may exist but are unstable, and not observed. This is the case of structures containing more than one mode at each point, which lead to structures of rhombs, and hexagons.

\section{\label{sec:conclusions}Conclusion}

	In this article, we extended a numerical scheme proposed by Christov
	\& Pontes~\cite{pontes2008} to investigate pattern formation
	modeled by the cubic Swift-Hohenberg equation  in two
	dimensions. The original scheme presents second order representation
	of all derivatives, strict implementation of the associated Lyapunov
	functional, rigid boundary conditions, and a semi-implicit
	assignment of the terms.
	The present work includes the quintic version of the Swift-Hohenberg
	equation and periodic boundary conditions for both the cubic
	and the quintic versions of the model. The scheme retains all
	characteristics of the original one, namely strict representation of
	the Lyapunov functional, unconditional stability, and second order
	representation of all derivatives. In addition, we also included a
	convergence analysis, new verification tests, and an initial
	evaluation of the effect of nonuniform forcings in the form of
	spatial ramps and of Gaussian distributions of the control parameter
	$\epsilon$. Among the verification tests, the convergence analysis
	confirmed the truly second-order accuracy of the scheme both in
	space and time and the existence of localized structures developed
	in the framework of the quintic equation even with nonuniform forcings. 
	The conducted tests confirmed the robustness of the developed
	tool for pursuing the investigation of the pattern formation
	through the parabolic Swift-Hohenberg equation presenting various
	nonlinear terms, including nonuniform forcings.

	Additionally, the numerical experiments conducted in this work suggest the
	existence of effects and the onset of patterns not addressed in the
	literature. Both questions are the object of a
	companion study~\cite{coelho2021stripe}, based on the present numerical framework.

\section{Acknowledgments}
	The authors thank FAPERJ (Research Support Foundation of the State of Rio de
	Janeiro) and CNPq (National Council for Scientific and Technological
	Development) for the financial support. Daniel Coelho acknowledges a
	fellowship from the Coordination for the Improvement of Higher Education
	Personnel-CAPES (Brazil). A FAPERJ Senior Researcher Fellowship is
	acknowledged by J. Pontes. The authors dedicate a special posthumous thanks
	to C. I. Christov, who performed great development regarding the numerical
	methods addressed in this paper.

\nocite{*}
\bibliography{main}

\end{document}